\documentclass{aa}

\usepackage{graphicx,url,twoopt,natbib}
\usepackage[varg]{txfonts}
\usepackage{subfigure} 
\usepackage{enumerate}
\usepackage{amsmath}
\usepackage{lineno}
\usepackage[colorlinks,
linkcolor=blue,
anchorcolor=blue,
citecolor=blue]{hyperref}

\bibpunct{(}{)}{;}{a}{}{,}

\newcommand{\Mdot}{\mbox{\,$\mathrm{M}_{\odot}$}}      
\newcommand{\Mdotrate}{\mbox{\,$\mathrm{M}_{\odot}/\mathrm{yr}$}}       

\begin{document}

\title{Dynamical Mass Loss at the End of TP-AGB stars}

\author{Yingzhen Cui\inst{1}\and Song Wang\inst{1,2}
\and Xiangcun Meng\inst{3}\and Jifeng Liu\inst{1,2,4,5}
\and Shuguo Ma\inst{1}\and Weitao Zhao\inst{6,7}}

\institute{Key Laboratory of Optical Astronomy, National Astronomical Observatories, Chinese Academy of Sciences, Beijing 100101, China\\e-mail: cuiyz@bao.ac.cn
\and Institute for Frontiers in Astronomy and Astrophysics, Beijing Normal University, Beijing 102206, China
\and International Centre of Supernovae (ICESUN), Yunnan Key Laboratory of Supernova Research, Yunnan Observatories, Chinese Academy of Sciences (CAS), Kunming 650216, China
\and School of Astronomy and Space Science, University of Chinese Academy of Sciences, Beijing 100049, China
\and New Cornerstone Science Laboratory, National Astronomical Observatories, Chinese Academy of Sciences, Beijing 100012, China
\and School of Physics, Henan Normal University, Xinxiang 453007, China 
\and Center for Theoretical Physics, Henan Normal University, Xinxiang 453007, China}

\abstract{The thermally pulsating asymptotic giant branch (TP-AGB) phase plays a key role in the evolution of low- to intermediate-mass stars, driving mass loss that influences their final stages and contributes to galactic chemical enrichment. However, the mechanisms behind mass loss, particularly at the end of AGB, are still not well understood.} {We aim to investigate the relationship between stellar parameters and envelope dynamics during the TP-AGB phase, evaluating whether dynamical instabilities in the envelope can act as a possible mass-loss mechanism.} {We use hydrodynamics method in MESA to simulate the dynamical pulsations and resulting mass loss during the TP-AGB phase of a star evolved from a 1.5\Mdot \ zero-age main sequence.} {Our simulations reproduce the dynamical pulsation behavior of stars during the TP-AGB phase, demonstrating that the envelope mass is a key factor governing pulsational properties. As the envelope mass decreases, both the pulsation period and radial amplitude increase,  consistent with observational trends. For 1.5\Mdot\ model, once the envelope mass declines to approximately 0.25\Mdot, the model enters a regime of violent pulsations, potentially ejecting the remaining envelope within a few hundred years.}{We suggest that the instability can act as the dominant mass-loss mechanism in the end of the TP-AGB phase, marking a rapid transitional stage toward the post-AGB phase.}


\titlerunning{Dynamical Mass Loss in TP-AGB Stars}
\authorrunning{Cui et al.}

\maketitle
\section{Introduction}
The thermally pulsating asymptotic giant branch (TP-AGB) represents a late and critical phase in the evolution of low- to intermediate-mass stars, during which the stars lose entire envelope and ultimately leave behind a white dwarf \citep{Willson2000ARA&A..38..573W}. This process is crucial for enriching the interstellar medium with heavy elements and dust, which contribute to the chemical and dynamic evolution of galaxies \citep{Marigo2001A&A...370..194M}. 

During the TP-AGB phase, stars undergo periodic helium shell flashes, known as thermal pulses, which are interspersed with quiescent hydrogen shell burning phases \citep{Schwarzschild1965ApJ...142..855S}. In addition to these thermal events, stars also exhibit hydrodynamical pulsations. According to the periods, these pulsations can be classified into three classes of variable stars in observation: semiregular (SR) variables, Mira variables, and OH/IR stars \citep{Zijlstra2006IAUS..234...55Z}. Semiregular variables exhibit pulsation periods of 50 to 150 days, small amplitudes, and irregular light curves. The pulsation periods range from 150 to 500 days in the Mira variables, where amplitudes exceed 2.5 magnitudes in the optical band. For OH/IR stars, the pulsation periods lengthen further, reaching 500 to 2000 days, and the energy distribution shifts to the infrared due to dust formation. 

Mass loss plays a pivotal role in the evolution and final fate of TP-AGB stars. \cite{Reimers1975MSRSL...8..369R} first derived an empirical mass loss formula for post-main-sequence stars. However, \cite{Renzini1981A&A....94..175R} argued that in order to explain the characteristics of planetary nebulae, a superwind phase is necessary near the tip of the AGB phase, during which the mass-loss rate can significantly exceed that predicted by Reimers' law. Subsequent studies revealed the correlation between mass loss and stellar pulsation in mass-losing AGB stars \citep{Schild1989MNRAS.240...63S,Wood1990ASPC...11..355W}: for pulsation periods up to approximately 500 days, the mass-loss rate increases exponentially with period, and beyond this threshold, the mass-loss rate remains roughly constant in the range of a few $10^{-5}$ to $10^{-4}$ \Mdotrate. Based on these results, \cite{Vassiliadis1993ApJ...413..641V} proposed a semi-empirical mass-loss prescription, which provides better agreement with observations when applied to stellar evolution models \citep{Willson2000ARA&A..38..573W,Rosenfield2014ApJ...790...22R,Lattanzio2016JPhCS.728b2002L,Hofner2018A&ARv..26....1H}.

In addition to the empirical prescriptions, efforts have also been made to understand the physical origin of mass loss. Currently, the dust-driven wind is the most popular mechanism to explain the mass loss in the TP-AGB stars \citep{Hofner2018A&ARv..26....1H,Decin2021ARA&A..59..337D,Hofner2022IAUS..366..165H}. In this mechanism, periodic large-amplitude pulsations shock the outer layers, lifting material to regions where temperatures are low enough for gas molecules to condense into dust grains. These grains interact with the stellar radiation field, absorbing and scattering light. The momentum transferring to the surrounding gas drives significant mass outflows. Based on the dust-driven wind theory, detailed grids of stellar wind models have been developed and show good agreement with observations in terms of mass-loss rates, wind velocities, and spectral energy distributions \citep{Maraston2005MNRAS.362..799M,Nowotny2011A&A...529A.129N,Nowotny2013A&A...552A..20N,Eriksson2014A&A...566A..95E,Bladh2015A&A...575A.105B,Bladh2019A&A...626A.100B}. 

Recent studies further emphasized the importance of mass loss during the very final AGB phases for understanding the spectra of post-AGB stars and planetary nebulae \citep{Tosi2022A&A...668A..22T,Dell'Agli2023MNRAS.526.5386D}. However, the mass loss at the end of AGB phase remains poorly understood. \cite{Olofsson2022A&A...665A..82O} investigated 17 OH/IR stars located in the Galactic Bulge and found that they can be divided into two distinct groups: 11 stars exhibit the typical characteristics of standard OH/IR stars, while the other 6 have detached envelopes. The significant difference between these two groups indicates the existence of a violent mass-loss phase of about $10^2$ years during the final stage of the AGB. However, because dust-driven winds are limited to mass-loss rates of about a few $10^{-5}$\Mdotrate \ \citep{Decin2019NatAs...3..408D}, they seem insufficient to explain such a rapid transition. Moreover, numerical simulations often encounter convergence issues near the end of the TP-AGB phase, which are generally considered as signatures of underlying physical instabilities \citep{Wagenhuber1994A&A...290..807W,Lau2012A&A...542A...1L}. Recently, \cite{Gautschy2023arXiv230311374G} suggested that these instabilities may take the form of strong pulsations, which could directly drive violent mass loss at the end of AGB. It is plausible that such instabilities are directly linked to the rapid transition phase as mentioned above. Therefore, a detailed investigation into the nature and consequences of these instabilities is essential for a complete understanding of TP-AGB evolution. 

In this work, we use one-dimensional hydrodynamic simulations to investigate the dynamical instability of the TP-AGB stars. This paper is organized as follows: In Sect.~2, we describe the input physics and computational framework. Sect.~3 presents the key results, highlighting the reproduction of Mira-like pulsations, the relationship between envelope mass and pulsation amplitude, and the process of the instability phase. In Sect.~4, we discuss the limitations of the current models, explore the physical implications of our findings, and give the conclusions. 

\section{Method}
We used the Modules for Experiments in Stellar Astrophysics (MESA) code \citep[version 10398;][]{Paxton2011ApJS,Paxton2013ApJS,Paxton2015ApJS,Paxton2018ApJS,Paxton2019ApJS} to evolve a 1.5\Mdot\ stellar model from the zero-age main sequence to the end of the TP-AGB phase, adopting a metallicity of 0.02 and a mixing-length parameter of 2. For stellar wind prescriptions, we adopted the Reimers wind formula with a coefficient of 0.5 during the red giant branch (RGB) phase and the Bl{\"o}cker wind formula with a coefficient of 0.05 during the AGB phase\footnote{In our 1.5\Mdot\ model, the star remains oxygen-rich throughout the entire AGB phase.}  \citep{Reimers1975MSRSL...8..369R,Bloecker1995A&A...297..727B}. At the tip of the RGB, the mass of the star is reduced to about 1.35\Mdot.

Then, we chose the point during the TP-AGB phase when the envelope mass has decreased to 0.64\Mdot\ as the starting point for the subsequent hydrodynamical simulation. This choice was made to ensure that the model is as close as possible to the end of the TP-AGB phase, while the envelope mass has not yet been excessively depleted, thereby allowing us to investigate how the dynamical properties of TP-AGB stars evolve with decreasing envelope mass toward the end of their evolution.

In the above simulations, we enforced the model to remain in hydrostatic equilibrium at all times. We then performed subsequent simulations using hydrodynamics module in MESA (see \cite{Paxton2015ApJS}). The default Eddington gray atmosphere in MESA was adopted as the outer boundary condition. To resolve the dynamical pulsations, which typically have periods of only a few hundred days, we restricted the time step to $10^{-3}$ years. 

The hydrodynamics module enables us to follow the velocity evolution of each layer in the stellar model. Once the velocity of the outer envelope exceeds the local escape speed, we regard this as the onset of mass ejection. In our simulations, we followed the approach of \cite{Clayton2017MNRAS.470.1788C}, removing envelope layers that exceed the local escape velocity in order to simulate the ejection of mass.

\section{Results}
\subsection{From hydrostatic models to dynamical simulations}
\begin{figure}
	\centering
	\resizebox{\hsize}{!}{\includegraphics{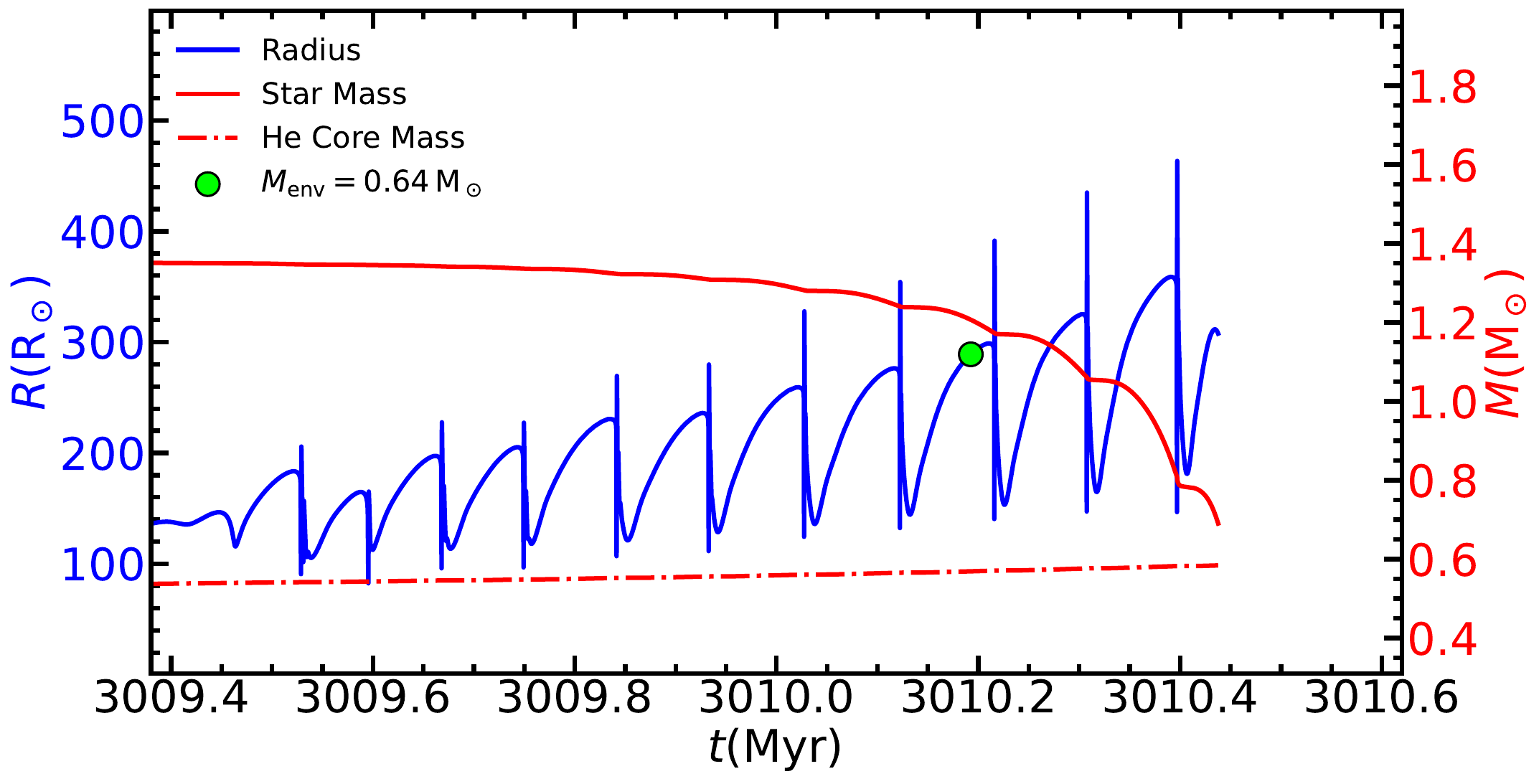}}
	\caption{The hydrostatic evolution during the TP-AGB phase. The blue line represents the model's radius, while the red lines indicate the total mass and helium core mass, respectively. The green point mark the selected 0.64\Mdot\ envelope model for subsequent hydrodynamical simulations.}
	\label{TPAGB}
\end{figure}

\begin{figure}
	\centering
	\resizebox{\hsize}{!}{\includegraphics{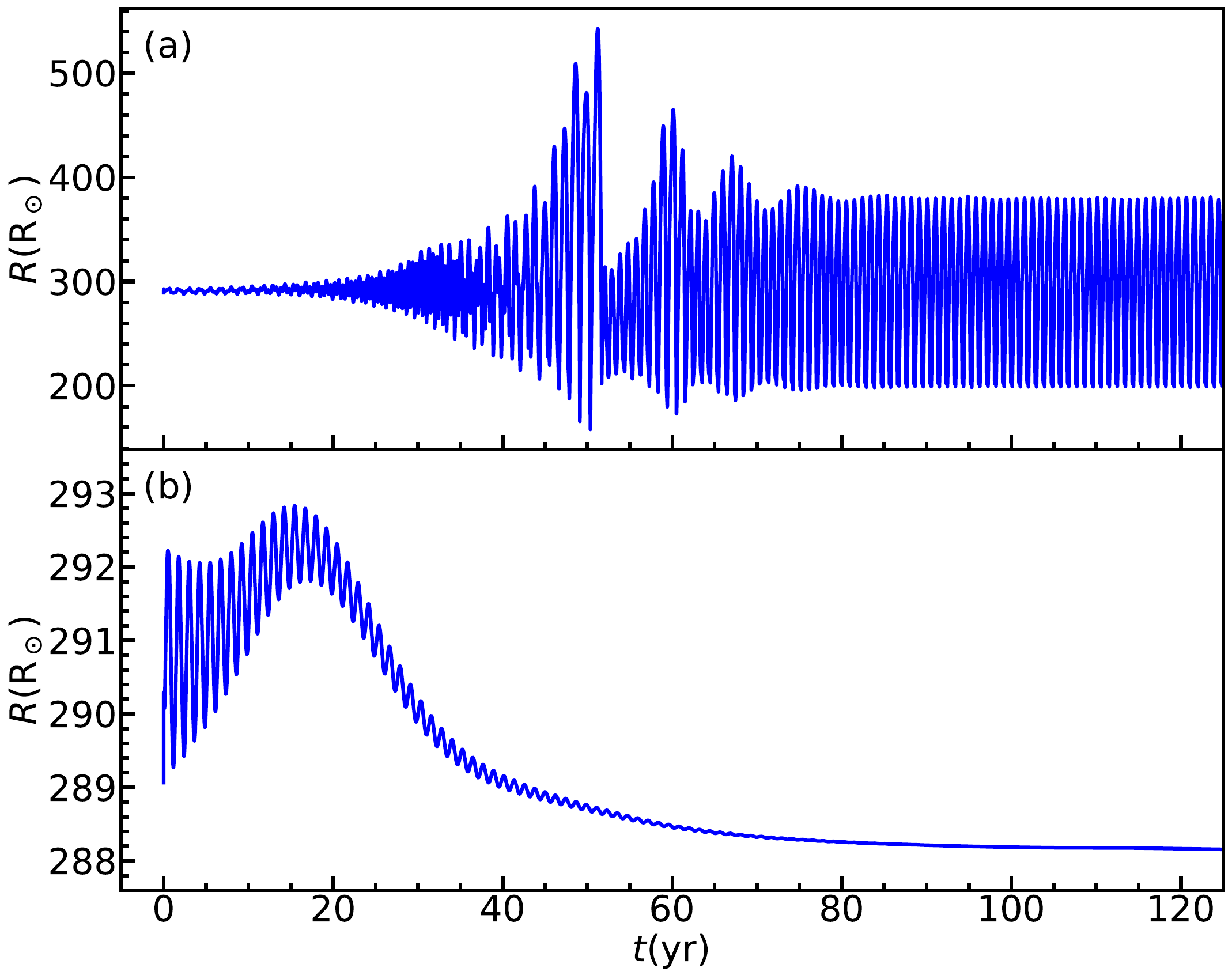}}
	\caption{The early stage of hydrodynamical simulations starting from a hydrostatic model with an envelope of 0.64\Mdot. For simplicity, we set $t = 0$ at the beginning of the simulation. Panel (a) uses a time step of $10^{-3}$ years, while panel (b) uses a time step of $10^{-2}$ years.}
	\label{dt}
\end{figure}

Fig.~\ref{TPAGB} shows the time evolution of the radius, core mass and total mass of the 1.5\Mdot\ model star discussed in the previous section, during the TP-AGB phase. The total mass gradually decreases as a consequence of the ongoing mass loss. The periodic variations in radius correspond to thermal pulses driven by unstable helium shell burning. The green point marks the model where the envelope mass has decreased to 0.64\Mdot, which we adopt as the initial condition for the hydrodynamical simulations.

Fig.~\ref{dt} shows the early stage of the hydrodynamical simulations for the 0.64\Mdot\ envelope model. In practice, to capture the dynamical effects, the timestep adopted in the simulations should be comparable to or smaller than the dynamical timescale of the model \citep{Heger1997A&A...327..224H,Yoon2010ApJ...717L..62Y}. To illustrate the impact of timestep choice, we present simulations with timesteps of $10^{-3}$ and $10^{-2}$ years, while the dynamical timescale of this model is of order $10^{-1}$ years. At the start of the hydrodynamical calculations, both cases exhibit weak pulsations. Then, the model with a timestep of $10^{-2}$ years quickly restore hydrostatic equilibrium whereas the simulation with a shorter timestep eventually develops dynamical pulsations. Based on this result, we adopt a timestep of $10^{-3}$ years in the following calculations to ensure proper resolution of the pulsation behavior. Besides the impact of the timestep, it is also important to note that a sufficiently long relaxation phase is required. As shown in Fig.~\ref{dt} (a), the model initially undergoes several cycles of pulsation growth before eventually settling into stable periodic pulsations.

\subsection{Evolution of the envelope instability}
\begin{figure}
	\centering
	\resizebox{\hsize}{!}{\includegraphics{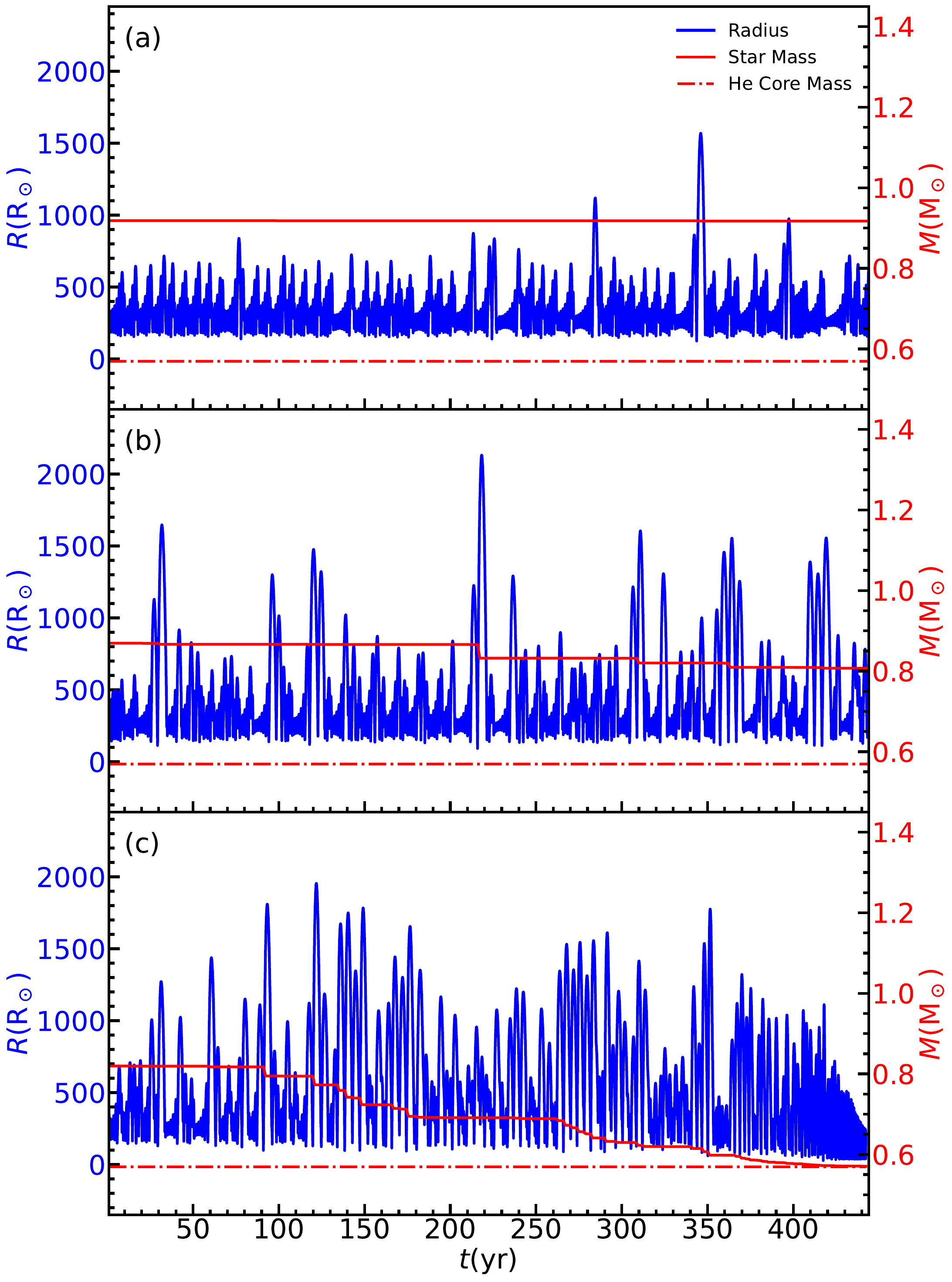}}
	\caption{Hydrodynamical evolution of the model with an initial envelope mass of 0.64\Mdot. The model first undergoes a relaxation phase to reach stable pulsations. At around $t \approx 120$ years, a constant mass-loss rate of $10^{-3}$\Mdotrate \ is introduced to reduce the envelope artificially.}
	\label{masslosstest}
\end{figure}

\begin{figure}
	\centering
	\resizebox{\hsize}{!}{\includegraphics{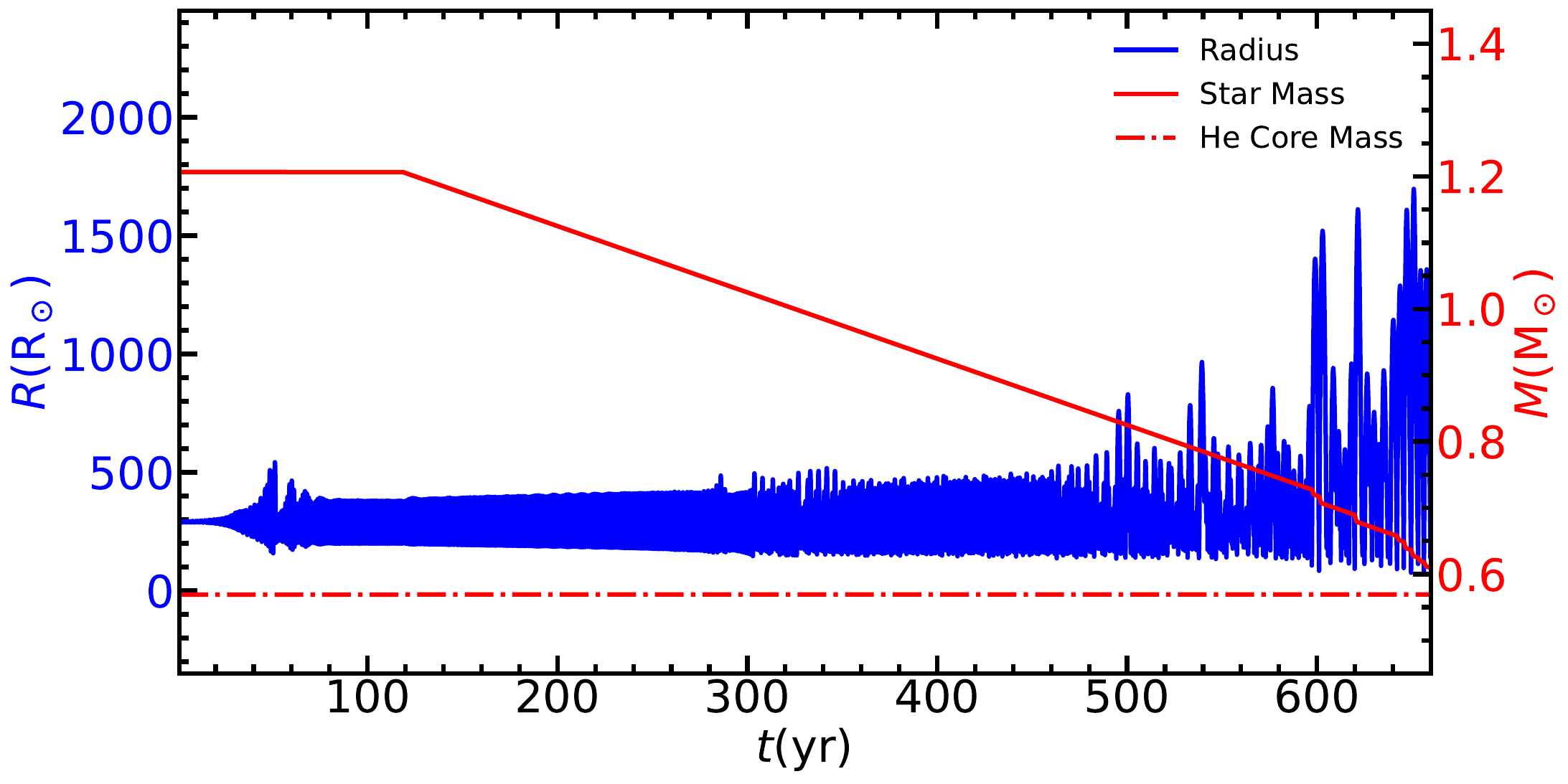}}
	\caption{Hydrodynamical evolution of models with different initial envelope masses. Panels (a), (b), and (c) correspond to the initial envelope masses of 0.36, 0.25, and 0.2\Mdot, respectively. The start of the simulation is set to $t = 0$ for all cases. }
	\label{ejection}
\end{figure}

\begin{figure}
	\centering
	\resizebox{\hsize}{!}{\includegraphics{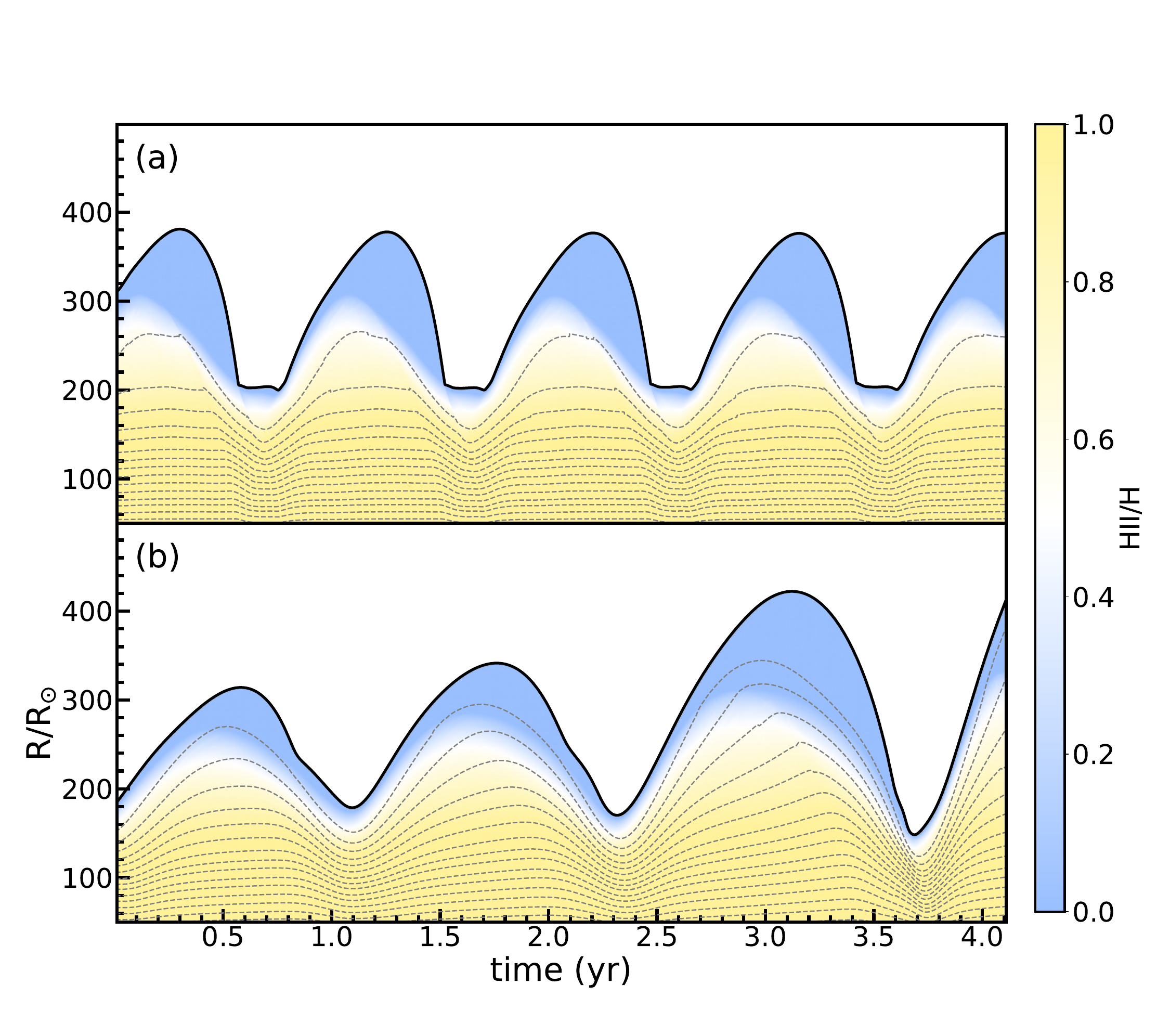}}
	\caption{Panels (a) and (b) show the evolution during the first four years after the onset of the dynamical simulation for models with masses of 0.64\Mdot\ and 0.25\Mdot, respectively. The black solid line indicates the stellar radius, while the gray dashed lines mark the radial positions that divide the envelope mass into 20 equal parts. The colorbar represents the relative fraction of ionized hydrogen.}
	\label{mechanism}
\end{figure}

\begin{figure}
	\centering
	\resizebox{\hsize}{!}{\includegraphics{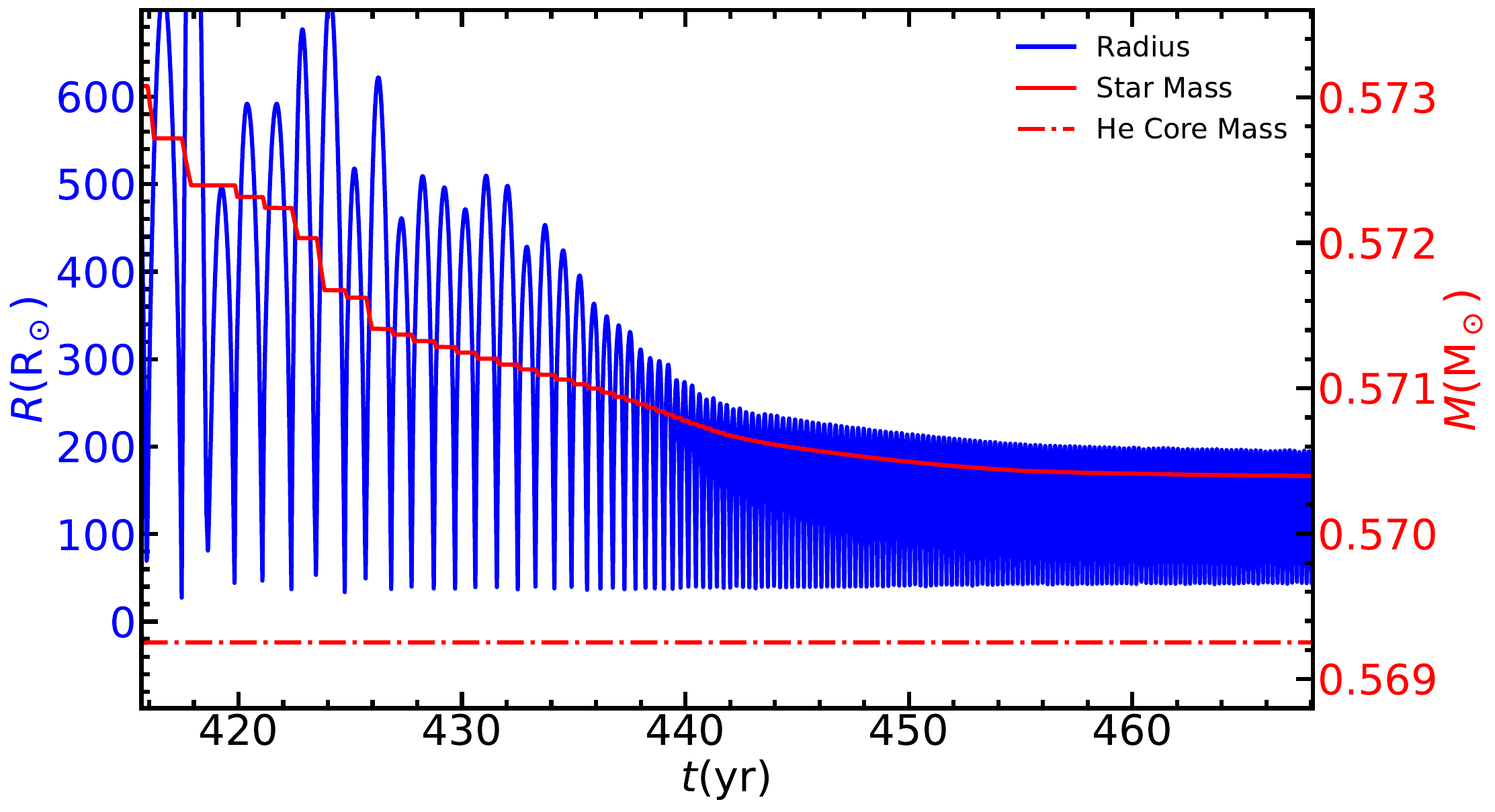}}
	\caption{Same evolutionary model as in Fig.~\ref{ejection}(c), but zoomed in to show the details during the final stages of the model's evolution. }
	\label{ejection_end}
\end{figure}

In this section, we aim to investigate how the dynamical pulsation properties of the model evolve along with stellar evolution, particularly as the star approaches the end of the TP-AGB phase. With the small time step of $10^{-3}$ years, it is not feasible to follow the entire TP-AGB evolution, which spans several $10^{5}$ years. A practical approach is to extract models at different evolutionary stages from the hydrostatic simulation and then use hydrodynamical simulations to reproduce their pulsation properties. Unfortunately, as the star evolves, the relaxation process from the hydrostatic to the dynamical state becomes increasingly violent, ultimately leading to numerical difficulties. An alternative approach is to shorten the TP-AGB evolutionary timescale. This can be achieved by adopting a mass-loss rate significantly higher than the typical Bl{\"o}cker wind prescription, since mass loss is the dominant factor controlling the pace of evolution \citep{Hofner2018A&ARv..26....1H}. This approach allows us to follow the changes in dynamical pulsation properties throughout the entire evolution via hydrodynamical simulations. 

Fig.~\ref{masslosstest} illustrates how the pulsation characteristics of the model evolve as the envelope mass gradually decreases. The model begins with an initial envelope mass of 0.64\Mdot. For the first 120 years, no mass loss is applied, and the model undergoes a relaxation process, eventually reaching a stable pulsation state. This phase corresponds to the behavior shown in panel (a) of the Fig.~\ref{dt}. After 120 years, we introduce an artificial stellar wind of $10^{-3}$\Mdotrate, leading to a steady decrease in the total mass of the model. This choice of stellar wind shortens the subsequent TP-AGB evolutionary timescale, while still keeping it long enough to allow us to trace the evolution of the pulsation characteristics. Until the stellar mass decreases to about 0.85\Mdot, corresponding to an envelope mass of approximately 0.3\Mdot, the pulsations remain relatively stable. However, as the envelope mass continues to decline, the pulsation amplitude begins to increase significantly. When the envelope mass drops below ~0.2\Mdot, the pulsation amplitude further increases, and the model begins to experience abrupt mass loss events. This occurs because the velocity of the outer envelope layers exceeds the local escape velocity, triggering mass ejection.

The above results were obtained under the presence of artificial continuous mass loss. Once the envelope becomes pulsationally unstable, the pulsation properties change rapidly as the envelope mass decreases. In this regime, it is difficult to disentangle whether the observed behavior is driven primarily by the imposed mass loss or by the intrinsic time development of the instability itself. To address this issue, we extracted models with envelope masses of 0.20, 0.25, and 0.30\Mdot\ from sequences described above, and then evolved them further in new hydrodynamical simulations without the $10^{-3}$\Mdotrate\ mass loss. In these subsequent simulations, any mass loss occurs self-consistently only when envelope material reaches velocities exceeding the local escape speed. The results are shown in Fig.~\ref{ejection}. For the 0.30\Mdot\ envelope model, pulsations were stronger than that in higher-mass cases but still insufficient to trigger a mass ejection. As the envelope mass decreases, mass ejection begins to occur during the pulsation cycles, as indicated by the sudden drop in stellar mass shown in the figure. In the case of the 0.2\Mdot\ envelope model, the ejection becomes so intense that the remaining envelope can be completely expelled within several hundred years. From our results, it appears that TP-AGB stars may experience a rapid mass-ejection phase in their final evolutionary stages. This ejection is driven by violent pulsations that accelerate a portion of the envelope beyond the escape velocity, resulting in an average mass-loss rate as high as $5\times10^{-4}$\Mdotrate, which can exceed that of a dust-driven wind by more than an order of magnitude. Such a process can lose nearly all of the remaining envelope within several hundred years, providing a natural explanation for the rapid transition phase at the end  of the TP-AGB stars suggested by \citet{Olofsson2022A&A...665A..82O}. 

The strengthening of pulsations with decreasing envelope mass can be understood in terms of the $\kappa$ mechanism \citep{Cui2022A&A...667A.154C,Gautschy2023arXiv230311374G}. Numerical results show that the $\kappa$ mechanism tends to excite pulsations in the partially ionized zones, whereas it acts as a damping process in other regions \citep{Cox1963ApJ...138..487C,Baker1965ApJ...142..868B}. Therefore, the larger the fraction of the envelope mass contained within the partially ionization zones, the more unstable the envelope becomes. To illustrate how this mechanism operates as the envelope mass decreases, Fig.~\ref{mechanism} shows the dynamical pulsations of models with envelope masses of 0.64 and 0.25\Mdot, respectively. The gray dashed lines divide the envelope into 20 equal-mass shells, and the color scale indicates the hydrogen ionization fraction where the white regions mark the locations of the partial ionization zones. As seen in the figure, although the initial pulsation amplitudes of the two models are comparable, the lower-envelope-mass model has more mass shells passing through the partial ionization zones. As a result, a larger fraction of the material participates in the $\kappa$ mechanism, making the envelope more unstable.

In Fig.~\ref{ejection_end}, we present a zoomed-in view of the final phase of the ejection process, corresponding to the model in Fig.~\ref{ejection} (c). As the envelope mass decreases to about 0.0024\Mdot, the pulsation amplitude starts to decline, with the mass-loss rate gradually decreasing until it eventually ceases, leaving about 0.0012\Mdot\ of envelope material. This decline occurs because the opacity, $\kappa$, also depends on envelope density, so the effect of the $\kappa$ mechanism is expected to weaken as the envelope mass decreases. 

\subsection{Periods of the pulsations}
\begin{figure}
	\centering
	\resizebox{\hsize}{!}{\includegraphics{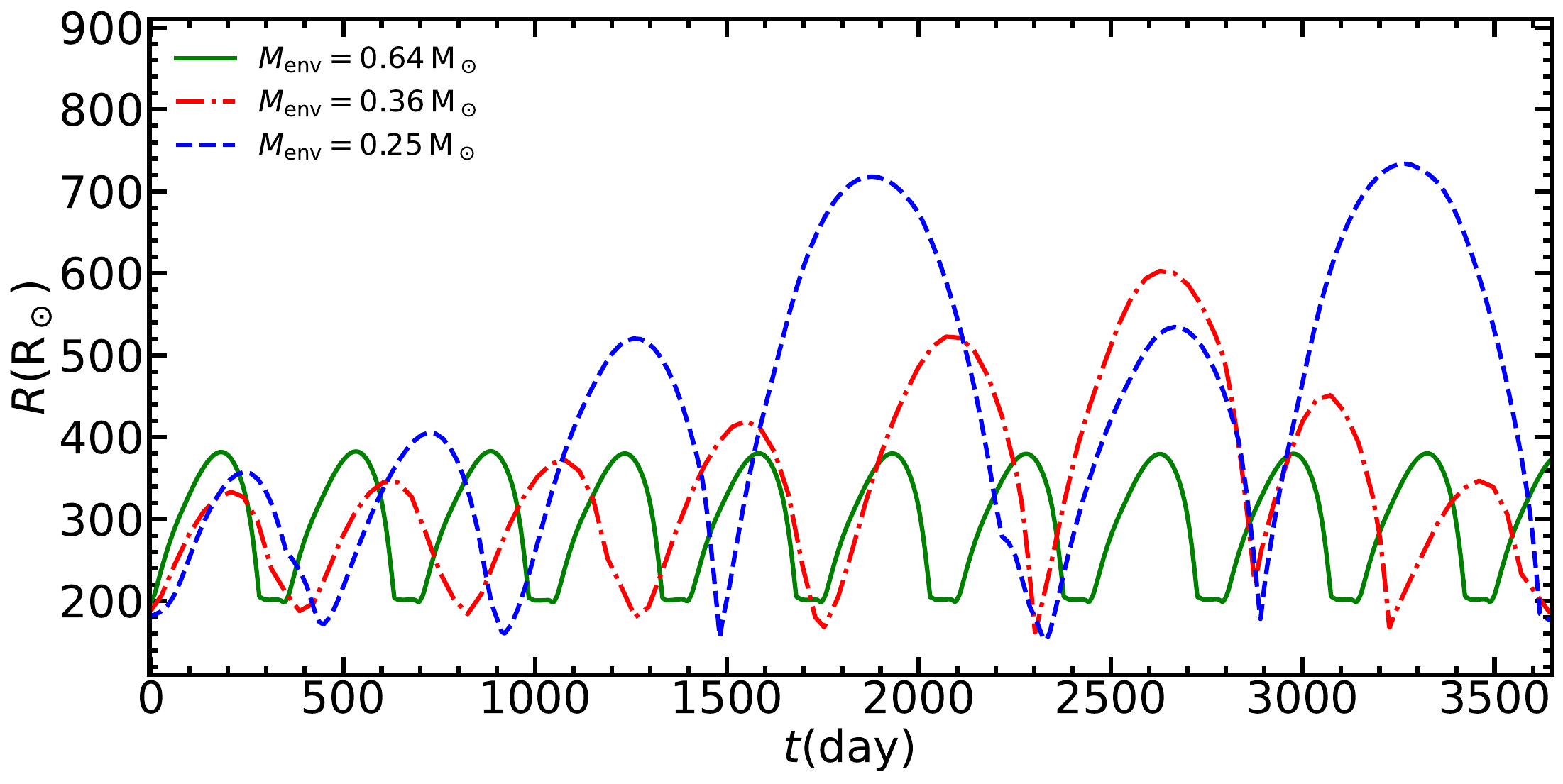}}
	\caption{Comparison of the pulsation properties among models with different envelope masses. For clarity, the time axis of each model has been shifted so that $t = 0$ corresponds to the minimum radius in its pulsation cycle. The envelope mass of each model is indicated in the figure.}
	\label{period}
\end{figure}
This section primarily illustrates how the pulsation period evolves with envelope mass. In Fig.~\ref{period}, we compare the pulsational properties of three models with envelope masses of 0.64, 0.36, and 0.25\Mdot. For clarity and consistency in comparison, we align all three models at the pulsation cycle where the radius reaches its minimum. Among them, the 0.64\Mdot\ model exhibits stable pulsations, with a pulsation period of approximately 350 days, consistent with the properties of observed Mira variables. In contrast, the lower-mass models display unstable pulsations, with both amplitude and period varying over time. Nevertheless, it is still evident that the pulsation period increases with decreasing envelope mass. In the case of the 0.25\Mdot\ envelope model, the average period can exceed 500 days, which may indicate that the star is transitioning from a Mira variable to an OH/IR star. Overall, our results indicate that, as the TP-AGB star evolves, it may sequentially display the pulsation period characteristics of Mira variables and OH/IR stars.

\section{Discussion and Conclusions}
In this work, we performed one-dimensional hydrodynamic simulations to investigate the dynamical pulsation behavior of stars during the TP-AGB phase. Our results show that, envelope mass is a key factor influencing the pulsation characteristics: as the envelope mass decreases, both the pulsation amplitude and period increase. This trend is consistent with the observed properties of Mira variables and OH/IR sources. Notably, we find that once the envelope mass drops below a certain threshold, approximately 0.25\Mdot\ for a star with an initial mass of 1.5 \Mdot, the pulsations intensify dramatically and induce violent mass loss. Such rapid mass ejection can remove the remaining envelope within a few hundred years, providing a possible explanation for the intense envelope ejection that may occur at the end of the TP-AGB phase. 

\cite{MillerBertolami2016A&A...588A..25M} presented detailed post-AGB evolutionary sequences and showed the dependence of the post-AGB evolution on the residual envelope mass at the end of the AGB. In that work, instabilities occurring during the late TP-AGB were also mentioned, but their physical origin and impact on the final envelope mass were not explicitly modeled. Our results suggest that dynamical pulsation-driven mass ejection may provide a possible physical mechanism contributing to the late-stage removal of the envelope. In the specific case of our 1.5\Mdot\ model, the instability does not lead to almost total removal of the envelope. In this sense, the general framework proposed by \citet{MillerBertolami2016A&A...588A..25M} is not expected to be significantly altered. Moreover, the use of grey opacities in our simulations make it difficult to accurately capture the evolution of the photospheric position once the envelope becomes very dilute, which currently prevents a reliable quantitatively assessing the implications of our results for the post-AGB evolution.

In addition to the 1.5\Mdot\ model discussed in detail in this work, we examined models with initial masses of 1.0\Mdot\ and 3.0\Mdot. We find that the instability threshold for violent mass loss depends strongly on the stellar mass: the 1.0\Mdot\ model only triggers strong ejection when the envelope mass decreases to 0.025\Mdot, whereas the 3.0\Mdot\ model becomes unstable at a much higher envelope mass of 0.6\Mdot. Furthermore, we also investigated the effect of metallicity. Using a 1.5\Mdot\ model with Z=0.001, we found that ejection only occurs once the envelope mass decreases to 0.035\Mdot\,, which indicates that lower metallicity tends to reduce the instability threshold. Based on the results of the above preliminary tests, a comprehensive grid of models appears to be necessary. In the future, we aim to present such a grid covering different parameters, which will allow us to systematically explore hydrodynamical behaviors of the TP-AGB envelopes.

Notably, similar pulsation-driven ejection processes may occur in other astrophysical contexts. One prominent example is red supergiants. \citet{Yoon2010ApJ...717L..62Y} demonstrated that red supergiants with initial masses above 19\Mdot\ can develop strong pulsational instabilities. They suggested that such pulsations could lead to dust-driven winds, potentially explaining the observed scarcity of supernovae from stars above this mass. Although their results did not show direct mass ejection driven by pulsations, such ejection can be expected in even more massive red supergiants. Another relevant case is common envelope systems. \citet{Clayton2017MNRAS.470.1788C} modeled the ejection process in common envelope systems by injecting energy into the envelope of a red giant. They found that under such conditions, the envelope undergoes strong pulsations accompanied by vigorous mass ejection. Combining the above studies, the pulsation-driven envelope ejection in our TP-AGB models may reflect a more general mass-loss mechanism, operating across a broad range of stellar environments.

\begin{acknowledgements}
We are grateful to the anonymous referee for his/her constructive comments that helps us to improve the manuscript greatly. This work was supported by National Natural Science Foundation of China (NSFC) under grant Nos. 12273057/12588202/11833002/12090042/12288102/12333008/11973080, the National Key Research and Development Program of China (NKRDPC) under grant number 2023YFA1607901/No.2021YFA1600403, the Strategic Priority Program of the Chinese Academy of Sciences under grant number XDB1160302/XDB1160303/XDB1160000, and science research grants from the China Manned Space Project. X.M. acknowledges support from International Centre of Supernovae, Yunnan Key Laboratory under grant number No. 202302AN360001, Yunnan Fundamental Research Projects Nos. 202401BC070007. J.L acknowledges the support from the New Cornerstone Science Foundation through the New Cornerstone Investigator Program and the XPLORER PRIZE. 
\end{acknowledgements}

\bibliographystyle{aa} 
\bibliography{ref}
\end{document}